\documentstyle[aps,prl,multicol,epsfig]{revtex}

\begin{document}

\draft

\author{Mario Castro$^{a),b)}$, Francisco Dom\'{\i}nguez-Adame$^{a)}$,
Angel S\'anchez$^{c)}$, and Tom\'as Rodr\'{\i}guez$^{d)}$}

\address{$^{a)}$GISC, Departamento de F\'{\i}sica de Materiales, 
Universidad Complutense, E-28040 Madrid, Spain\\
$^{b)}$Universidad Pontificia de Comillas, E-28015 Madrid, Spain\\
$^{c)}$GISC, Departamento de Matem\'{a}ticas, Universidad Carlos III de Madrid, 
E-28911 Legan\'es, Madrid, Spain\\
$^{d)}$Departamento de Tecnolog\'{\i}a Electr\'onica, E.\ T.\ S.\ I.\ 
Telecomunicaci\'{o}n, Universidad Polit\'ecnica, E-28040 Madrid, Spain}

\title{Model for crystallization kinetics: Deviations from
Kolmogorov-Johnson-Mehl-Avrami kinetics} 

\maketitle

\begin{abstract}

We propose a simple and versatile model to understand the  deviations from the
well-known Kolmogorov-Johnson-Mehl-Avrami kinetics theory found in metal
recrystallization and amorphous semiconductor crystallization.  We analyze the
kinetics of the transformation and the grain size distribution of the product
material, finding a good overall agreement between our model and available
experimental data. The information so obtained could help to relate the
mentioned experimental deviations due to preexisting anisotropy along some
regions, to certain degree of crystallinity of the amorphous phases during
deposition, or more generally to impurities or roughness of the substrate.

\end{abstract}


\begin{multicols}{2} 

\narrowtext

The interest in thin film transistors made of polycrystalline silicon and
silicon-germanium has been driven by the technological development of active
matrix-addressed flat-panel displays~\cite{Im} and thin film solar
cells~\cite{Bergmann}. In this context, the capability to engineer the size and
geometry of grains becomes crucial to design materials with the required
properties. Crystallization of these materials takes place by nucleation and
growth mechanisms: Nucleation starts with the appearance of small atom clusters
({\em embryos\/}). At a certain fixed temperature, embryos with sizes greater
than a critical one become stable nuclei; otherwise, they shrink and eventually
they vanish. Such a critical radius arises from the competition between surface
tension and free energy density difference between amorphous and crystalline
phases (which favours the increasing of grain volume)  yielding an energy
barrier that has to be overcome to build up a critical nucleus. Surviving
nuclei grow by incorporation of neighboring atoms, yielding a moving boundary
with temperature dependent velocity that gradually covers the untransformed
phase. Growth ceases when growing grains impinge upon each other, forming a
grain boundary. The final product consists of regions separated by grain
boundaries. This simple picture has, however, two problems: On the one hand,
this theory of nucleation and growth predicts an energy barrier far from the
experimental value so nucleation would hardly be probable at available
annealing temperatures~\cite{Doherty}. On the other hand, it is known that in
crystallization of Si over SiO$_2$ substrates, nucleation develops in the
Si/SiO$_2$ interface due to inhomogeneities or impurities that catalyze the
transformation~\cite{Silicon}. Therefore, a theory of homogeneous nucleation
and growth is not entirely applicable to the referred experiments.

The transformation kinetics is also problematic. It is generally accepted that
the fraction of transformed material during crystallization, $X(t)$, obeys the
Kolmogorov-Johnson-Mehl-Avrami (KJMA) model~\cite{KJMA}, according to which 
$X(t)=1-\exp(-At^m),$ where $A$ is a nucleation- and growth-rate dependent
constant and $m$ is an exponent characteristic of the experimental conditions.
Two well-defined limits have been extensively discussed in the literature: When
all the nuclei are present and begin to grow at the beginning of the
transformation, the KJMA exponent, $m$, is equal to $2$ (in two dimensions),
and the nucleation is termed {\em site saturation\/}. The product
microstructure is tesselated by the so-called Voronoi polygons (or Wigner-Seitz
cells). On the contrary, when new nuclei appear at every step of the
transformation, $m=3$ and the process is named {\em continuous nucleation\/}.
Plots of $\log[-\log(1-X)]$ against $\log(t)$ should be straight lines of slope
$m$, called KJMA plots. The validity of the KJMA theory has been questioned in
the last few years \cite{Baram}, and subsequently several papers have been
devoted to check it in different ways~\cite{Sessa,Fanfoni,Siclen}. However,
those theoretical results still leave some open questions: For example, an
exponent between $2$ and $3$ is experimentally obtained in two dimensions, the
KJMA plots from experimental data do not fit a straight line in some
cases~\cite{Price}, and the connection between geometrical properties (grain
size distributions) and the KJMA exponent is not clear.

In this letter, we show that these questions may be answered by assuming  that
nucleation is heterogeneous, not in a phenomenological way as in other proposed
models~\cite{Thompson}, but sticking to the basic ideas due to
Cahn~\cite{Cahn1,Cahn2} and Beck~\cite{Beck}: The material is not perfectly
homogeneous but contains regions with some extra energy (regions with some
order produced during deposition, or substrate impurities) at which nucleation
is more probable. Accordingly, we introduce a computational model consisting of
several simple irreversible rules, with the additional advantage that it
describes simultaneously space and time evolution.  Furthermore, it allows us
to average over a large number of realizations in very short  computational
times as compared to other computer models (see the recent review by
Rollett~\cite{RolletReview} for an overview of simulation models of
recrystallization).

The model is defined on a two-dimensional lattice (square and triangular
lattices were employed) with periodic boundary conditions. Every lattice site
(or node) $\mathbf{x}$ belongs to a certain grain or state, $q({\mathbf
x},t)=0,1,2,$\ldots, the state $0$ being that of an untransformed region. The
lattice spacing is a typical length scale related to the available experimental
resolution. Following the idea that the amorphous phase has random regions in
which nucleation is favored, we choose a fraction $c$ of the total lattice
sites to be able to nucleate. We term these energetically favorable sites {\em
potential\/} nuclei. These potential sites may be interpreted as random sites
on a region where order is present, not just an isolated critical cluster. 
Initially $q({\mathbf x},0)=0$ for all lattice sites ${\mathbf x}$ and the
system evolves by parallel updating according to the following rules: i) A
transformed site remains in the same state [$q(x,t+\Delta t)=q(x,t)\neq 0$];
ii) An untransformed {\em potential\/} site may become a new non-existing state
(i.e., crystallizes) with probability $n$ (nucleation probability), if and only
if there are no transformed nearest neighbors around it; iii) An untransformed
site (including potential sites) transforms to a already existing transformed
state with probability $g$ (growth probability), if and only if there is at
least one transformed site on its neighborhood. The new state is randomly
chosen among the neighboring grain states.

\begin{figure}

\hspace*{.3in}
\epsfig{figure=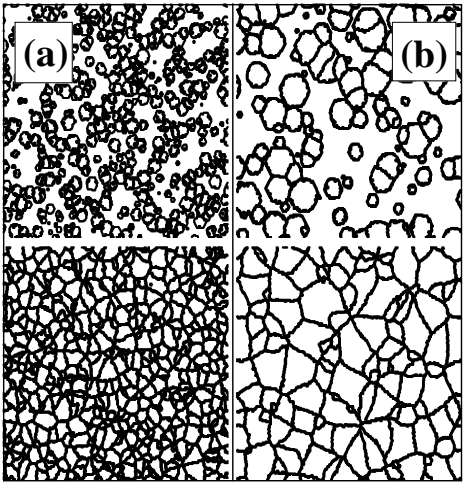,width=2.4in}
\vspace*{.1cm}

\caption{Computer simulation obtained for two stages of the transformation
process  on a $250\times 250$ triangular lattice with (a) $c=1$ (homogeneous
nucleation), $n=0.001$ and $g=0.8$ (total run time, 1 s.);  (b) $c=0.1$,
$n=0.001$ and $g=0.8$ (total run time, 1 s.).}
\label{micro}
\end{figure}

For the model parameters, we expect a functional form  $n\sim e^{-E_{n}/k_BT}$
and $g\sim e^{-E_{g}/k_BT}$, where $E_n$ and $E_g$ are the energy barriers of
nucleation and growth respectively. Hence, temperature is implicit in the
definition of $n$ and $g$. Figure~\ref{micro} shows the microstructure at two
different stages for two different sets of parameters. As we are interested in
this paper in how different nucleation conditions yield different KJMA
exponents and different microstructures for isothermal experiments, we define a
characteristic time $\tau$ as the time that a grain  needs to increase its size
by one lattice site, and consequently we can put $g=1$. The simulation time
step is therefore this characteristic time $\tau$.

We have simulated $1000\times1000$ triangular and square lattices and averaged 
the outcome of $50$ different realizations for each choice of parameters 
(characteristic simulation times are about $15$ to $45$ minutes in a Pentium II
personal computer). The main results are the following: If  $c\lesssim 1$, then
most sites are potential sites, so new grains are able to nucleate at every
stage of the transformation (continuous nucleation). On the contrary, when
$c\ll 1$, and $n\lesssim 1$, every potential site nucleates at the early stages
of the process (site saturation). Obviously, intermediate values yield a mixed
behaviour. Interestingly, the model parameters tune the KJMA exponent between
$2$ and $3$. It is important to note that for small values of $c$, which would
in principle mean  that growth is by site saturation, low values of $n$ (large
energy  barriers for nucleation) lead to $m\simeq 3$, as in continuous
nucleation.

\begin{figure}

\hspace*{.3in}
\epsfig{figure=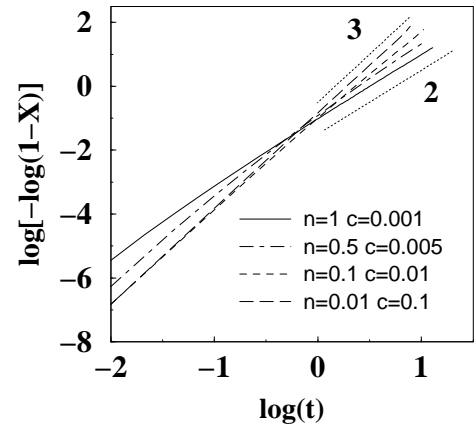,width=2.4in}
\vspace*{.1cm}

\caption{KJMA plots for different sets of parameters. The dotted lines
represent the theoretical slopes $2$ and $3$.}
\label{histo}
\end{figure}
Other forms of experimental behavior lead to the occurrence of non-straight
KJMA plots. We argue that this fact may be due, on the one hand, to the decay
of the nucleation rate when $n\ll 1$, because some potential sites are
overlapped by already growing grains; and on the other hand, when the potential
site concentration is $c\ll 1$, the grains grow independently for times lower
than a characteristic impingement time, proportional to the mean grain distance
$1/c^{1/2}$. Figure~\ref{histo} shows this fact for several choices of
parameters $n$ and $c$. Note that when $n\lesssim 1$, the potential sites
nucleate during the earlier stages of the transformation, so the mentioned
overlapping of potential sites cannot be the cause of the {\em bending} of the
KJMA plots. Therefore, we must conclude that heterogeneous nucleation is not
the unique cause of the unexpected bending of the KJMA plots, as $m$ may be
affected by anisotropies or preferential crystalline directions yielding growth
or nucleation rates that may change locally throughout the material. This
agrees with the fact  $m$ is not a reliable guide to characterize the
morphology of the evolving grains~\cite{Vandermeer}. 

\begin{figure}

\hspace*{.3in}
\epsfig{figure=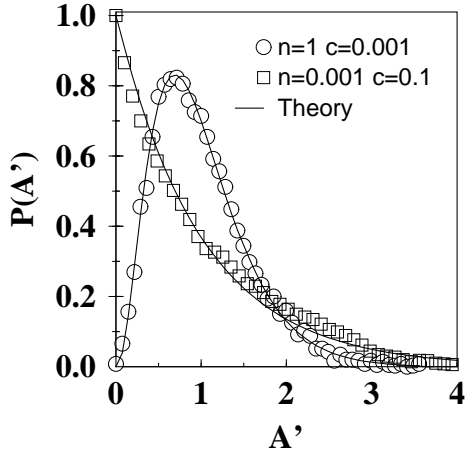,width=2.4in}
\vspace*{.1cm}

\caption{Histograms of the grain size distribution for different sets of 
parameters.}
\label{histo_2}
\end{figure}

As we have pointed out, our model provides information about microstructure,
i.e., number of grains, mean grain area, grain  size distribution, and so on.
For site saturation, Weire {\em et al.}\ proposed a {\em phenomenological\/}
expression for grain size distributions~\cite{Weire}:
$P(A^{\prime})=(A^\prime)^{\alpha-1}\alpha^\alpha e^{-\alpha
A^\prime}/\Gamma(\alpha)$, where $\alpha\approx 3.65$ and $A^\prime=A/\bar{A}$
is the reduced area. The mean area $\bar{A}$ changes from one process to
another, but the normalized distribution is the same for all. Analogously, in
the case of continuous nucleation, a simple expression has been
proposed~\cite{Mulheran}: $P(A^{\prime})=e^{-A^\prime}$. Figure \ref{histo_2}
shows the good agreement between the simulations of our model and these
theoretical predictions. For intermediate-ranging parameters, a continuous 
evolution is obtained from site saturation to continuous nucleation grain size
distributions. We thus have two elements of comparison between our model and
experimental results: the KJMA exponent, $m$, and the grain size distribution
$P(A^\prime)$.

In conclusion, we have presented a simple lattice model for crystallization 
which sheds light on the possible causes of the experimental deviations from
the  KJMA theory. Thus, preexisting inhomogeneities in the initial state, such
as regions with a lesser degree of disorder or impurities, dramatically change
the product structure and the time development of the crystalline phase. One of
the remarkable points of our model is its versatility, so other ingredients can
be simply added to the model rules. We postpone the detailed study of
heterogeneous growth or preferential directions to further research. The main
conclusion of this work is that the KJMA exponent is not enough to understand
and to characterize the crystallization mode in a specific experiment: Indeed,
we have shown that conditions close to site saturation and continuous
nucleation give rise to very similar values of $m$. Therefore, studies of the
grain size distribution are indispensable to identify correctly the
crystallization mode. We stress that the model rules are physically meaningful
(alternative proposals can be found in Ref.~\cite{Hesselbarth}, but are far
from being physical because they depend strongly on the lattice geometry and
the site interactions), and lead to experimentally verifiable predictions. Due
to its versatility and short simulation times, it is an easy to reproduce,
good, and nonexpensive testbed for the design of materials and  structures with
tailored grain size or shape properties.

The authors wish to thank A.\ Rodr\'{\i}guez, J.\ Olivares, and C.\ Ballesteros
for stimulating discussion on experimental issues, and E.\ Maci\'a for helpful
comments. This work has been supported by CAM under project 07N/0034/98 (M.\
C.\ and F.\ D.-A.), by DGES under project PB96-0119 (A.\ S.), and by MAT96-0438
(T.\ R.).

\end{multicols}
\end{document}